\documentclass[aip,pof,showpacs,reprint,onecolumn]{revtex4-1}

\usepackage[english]{babel}
\usepackage{epsfig}
\usepackage{amsmath}
\usepackage{comment}
\usepackage{graphicx}
\usepackage{nicefrac}

\newcommand{\bs}{\boldsymbol}
\newcommand{\mat}{\mathrm}


\begin{document}

\author{Michael Wilczek}
\email{mwilczek@uni-muenster.de}
\affiliation{Institute for Theoretical Physics, University of M\"unster, Wilhelm-Klemm-Str. 9, 48149 M\"unster, Germany}

\author{Benjamin Kadoch}
\affiliation{M2P2-CNRS, Aix-Marseille Universit\'e \& Ecole Centrale de Marseille, 38 Rue Joliot-Curie, 13451 Marseille Cedex 20, France\\Present address: IUSTI-CNRS, Polytech Marseille, Aix-Marseille Universit\'e}

\author{Kai Schneider}
\affiliation{M2P2-CNRS \& CMI, Aix-Marseille Universit\'e, 39 Rue Joliot-Curie, 13453 Marseille Cedex 13, France}

\author{Rudolf Friedrich}
\affiliation{Institute for Theoretical Physics, University of M\"unster, Wilhelm-Klemm-Str. 9, 48149 M\"unster, Germany}

\author{Marie Farge}
\affiliation{LMD-CNRS, Ecole Normale Sup\'erieure, 24 Rue Lhomond, 75231 Paris Cedex 5, France}

\title{Conditional vorticity budget of coherent and incoherent flow contributions in fully developed homogeneous isotropic turbulence}

\begin{abstract}
We investigate the conditional vorticity budget of fully developed three-dimensional homogeneous isotropic turbulence with respect to coherent and incoherent flow contributions. The Coherent Vorticity Extraction based on orthogonal wavelets allows to decompose the vorticity field into coherent and incoherent contributions, of which the latter are noise-like. The impact of the vortex structures observed in fully developed turbulence on statistical balance equations is quantified considering the conditional vorticity budget. The connection between the basic structures present in the flow and their statistical implications is thereby assessed. The results are compared to those obtained for large- and small-scale contributions using a Fourier decomposition, which reveals pronounced differences.
\end{abstract}

\maketitle

\section{Introduction}

The problem of turbulence remains a paradigm for non-equilibrium statistical mechanics. The challenge comes from the spatio-temporal complexity which calls for a statistical description of the phenomenon, moreover the presence of coherent structures induces strong statistical correlations. For example, already the single-point vorticity statistics displays a highly non-Gaussian shape, which indicates pronounced spatial correlations of the vorticity field. Visualizations of vorticity in fully developed turbulent flows show that these correlations become manifest in the form of slender vortex tubes, which form a complex entangled global structure \cite{Sigg81,she90nat,Vincent91,douady91prl,Jimenez93,Ishihara}. Hence, one of the most interesting problems in turbulence research is to understand the relation between the coherent structures and their implications for statistical properties of the flow.\par
In this context it is particularly interesting to study dynamical rather than purely kinematic statistical relations. Maybe the most fundamental dynamical balance equation related to vorticity is the balance of enstrophy production and dissipation. Deriving an equation for the vorticity probability density function (PDF) is  even more informative and allows to study the conditional budget of enstrophy production and dissipation, or equivalently the balance of conditional vortex stretching and diffusion, where the ordinary budget equation is contained as a special case. This budget equation was introduced by Novikov and has been studied in a number of publications \cite{novikov93jfr,novikov94mpl,mui96pre}. The conditional vorticity budget allows to quantify vortex stretching and vorticity diffusion as a function of vorticity magnitude and hence to statistically discriminate strong vorticity regions in the flow from weak ones. One, however, would like to go one step beyond and disentangle the influence of coherent and incoherent flow contributions on these statistical quantities, which is possible with the help of the orthogonal wavelet decomposition. Farge et al. \cite{Farge_1999,Farge_2001} proposed a method, called Coherent Vorticity Extraction (CVE), to extract the coherent structures out of turbulent flows. This technique is based on a denoising of vorticity in wavelet space. It was shown that the CVE method is more efficient than Fourier filtering \cite{Farge_2003} and that fewer wavelet coefficients are necessary to reconstruct the coherent structures with increasing Reynolds number \cite{Okamoto_2007}, which means that the CVE method becomes more attractive as the flow becomes more intermittent. This method was applied to study the vortical structures in sheared and rotating turbulence \cite{Jacobitz_2008} and mixing layers \cite{Schneider_JFM_2005}. For all investigated flows it was shown that the coherent vortices are well represented with few wavelet coefficients and the statistics of the remaining background flow exhibits more Gaussian-like behavior. Moreover, the Coherent Vorticity Simulation (CVS), which models turbulent flows by considering only the time evolution of the coherent contribution while neglecting the incoherent one to model turbulent dissipation, was introduced by Farge and co-workers \cite{Farge_1999,Farge_2001_2}. The wavelet and Fourier nonlinear filtering methods were compared recently \cite{Yoshimatsu_2010}.\par
The aim of the present article is twofold. First, we combine the statistical analysis of the vorticity field with the Coherent Vorticity Extraction in order to obtain new insights on the coherent and incoherent contributions to the statistics. This gives a characterization of the statistical impact of coherent structures. Second, the results also serve as a benchmark to characterize the performance of the wavelet decomposition. For example, aiming at coherent vorticity simulations, it is especially desirable that the coherent contributions are driving the nonlinear dynamics. One of the most simple checks consequently is to study their contributions to dynamical budget equations such as the enstrophy budget. For comparison we include an analysis of low and high pass Fourier-filtered vorticity fields using the same number of degrees of freedom. This is an interesting investigation in its own right as it yields a characterization of large- and small-scale contributions and their interaction.\par
The remainder of this article is structured as follows. We first review the theoretical background for the conditional vorticity budget, the orthogonal wavelet decomposition and the principle of the Coherent Vorticity Extraction. Then, after summarizing some technical details on the direct numerical simulations performed for this work, we will present and discuss the numerically obtained results, before we conclude.

\section{PDF Equation and Conditional Vorticity Budget}
The dynamics of incompressible flows can be described in terms of the vorticity field $\bs \omega(\bs x,t) = \nabla \times \bs u(\bs x,t)$, defined as the curl of the velocity field. Its evolution equation takes the form
\begin{equation}\label{eq:vorticity}
  \frac{\partial}{\partial t} \bs\omega + \bs u \cdot \nabla \bs\omega = \mathrm{S} \bs\omega + \nu \Delta \bs \omega + \nabla \times \bs F ,
\end{equation}
where $\mathrm{S}(\bs x,t)=\frac{1}{2}\left[ \nabla\bs u(\bs x,t) +(\nabla \bs u(\bs x,t))^T \right]$ denotes the rate-of-strain tensor, $\nu$ denotes the kinematic viscosity, and $\bs F(\bs x,t)$ represents an external large-scale forcing applied to the flow in order to maintain a statistically stationary flow. As we are dealing with incompressible flows, we additionally have
\begin{equation}
  \nabla \cdot \bs u = 0.
\end{equation}
If one is now interested in the single-point statistics of the vorticity, a comprehensive characterization can be obtained by studying the evolution equation of the vorticity probability density function. The vorticity PDF can be introduced as an ensemble average ($\langle \cdot \rangle$) over the fine-grained PDF (the delta distribution) according to \cite{lundgren67pof}
\begin{equation}
  f(\bs \Omega;\bs x,t) = \big\langle \delta(\bs \omega(\bs x,t)-\bs \Omega) \big\rangle ,
\end{equation}
by which we have introduced the sample space variable $\bs \Omega$. In general this PDF will be a function of space and time, but for statistically homogeneous and stationary turbulence the PDF becomes independent of these variables. For isotropic turbulence the PDF only depends on the magnitude $\Omega$ of the vorticity, such that the PDF of the vorticity vector, $f(\bs \Omega)$, can be expressed in terms of the PDF of the magnitude of the vorticity, $\check{f}(\Omega)$, according to
\begin{equation}\label{eq:pdfisotropic}
  \check f(\Omega) = 4\pi\Omega^2f(\bs \Omega).
\end{equation}
Consequently the single-point statistics of statistically stationary homogeneous isotropic turbulence is fully characterized by the PDF of the vorticity magnitude only.\par
Regarding the evolution of this quantity, standard PDF methods allow to derive an evolution equation for the single-point vorticity PDF of homogeneous turbulence which takes the form \cite{lundgren67pof,novikov68sdp,pope00book,wilczek09pre}
\begin{equation}\label{eq:f1continuity}
  \frac{\partial}{\partial t} f(\bs \Omega;t) = -\nabla_{\bs \Omega}\cdot \big[ \big\langle \mat S\bs \omega + \nu \Delta \bs \omega + \nabla \times \bs F \big| \bs \Omega \big\rangle f(\bs \Omega;t) \big] .
\end{equation}
The temporal change of the vorticity PDF is accordingly given by the divergence of the conditionally averaged right-hand side of equation \eqref{eq:vorticity} times the vorticity PDF, such that this equation takes the form of a continuity equation (or conservation law) for the probability density. That means, the closure problem of turbulence in this formulation appears in terms of the unknown conditional averages of the vortex stretching term, the diffusive term and the external forcing. Once these functions are known, equation \eqref{eq:f1continuity} is closed. For isotropic turbulence these terms can be simplified further as they have to take the form
\begin{align}
  \big\langle \mat{S}\bs \omega \big | \bs \Omega \big\rangle = s(\Omega,t) \, \widehat {\bs \Omega} &\qquad s(\Omega,t) = \big\langle \widehat {\bs \omega} \cdot \mat{S}\bs \omega \big |\Omega \big\rangle , \label{eq:vortexstretching} \\
\big\langle \nu \Delta \bs \omega \big | \bs \Omega \big\rangle = d(\Omega,t) \, \widehat {\bs \Omega} &\qquad d(\Omega,t) = \big\langle \nu \widehat {\bs \omega} \cdot \Delta \bs \omega \big | \Omega \big\rangle , \label{eq:diffusive} \\
\big\langle \nabla \times \bs F \big | \bs \Omega \big\rangle = e(\Omega,t) \, \widehat {\bs \Omega} &\qquad e(\Omega,t) = \big\langle \widehat {\bs \omega} \cdot (\nabla \times \bs F) \big | \Omega \big\rangle , \label{eq:forcing}
\end{align}
in order to maintain invariance under arbitrary rotations (and reflections). This shows that these terms can be characterized by scalar-valued functions depending only on the magnitude of vorticity, which can be obtained by projecting the vectorial conditional averages on the direction of the vorticity, $\widehat{\bs \omega}=\bs \omega/\omega$, where $\omega = \lVert\bs \omega\rVert$. For a more detailed account on the exploitation of statistical symmetries we would like to refer the reader to Wilczek et al. \cite{wilczek11jfm}. If we insert these expressions together with relation \eqref{eq:pdfisotropic} into the PDF equation \eqref{eq:f1continuity}, we obtain the transport equation for the PDF of the magnitude of vorticity
\begin{equation}\label{eq:pdfiso}
\frac{\partial}{\partial t} \check f(\Omega;t)=-\frac{\partial}{\partial \Omega} \left[ s(\Omega,t)+d(\Omega,t)+e(\Omega,t) \right] \check f(\Omega;t) ,
\end{equation}
by which the problem becomes eventually one-dimensional; the temporal evolution of the vorticity PDF may fully be characterized by knowing the functions $s$, $d$ and $e$, which are related to the vortex stretching term, the diffusive term and the external forcing term, respectively. When we now consider statistically stationary turbulence, this equation implies
\begin{equation}
  s(\Omega)+d(\Omega)+e(\Omega) = 0
\end{equation}
as the probability current for this type of stationary one-dimensional problem has to vanish. Furthermore, it has been shown in a number of publications \cite{novikov93jfr,novikov94mpl,wilczek09pre} that the conditional vortex stretching and diffusive term balance at sufficiently high Reynolds numbers. Compared to those terms, the external forcing has a negligible effect, such that we obtain the approximation
\begin{equation}
  s(\Omega)+d(\Omega) \approx 0,
\end{equation}
or equivalently
\begin{equation}\label{eq:balance}
  \left[ s(\Omega)+d(\Omega)\right] \widehat{\bs \Omega} = \big\langle \mat{S}\bs \omega \big | \bs \Omega \big\rangle + \big\langle \nu \Delta \bs \omega \big | \bs \Omega \big\rangle \approx \bs 0.
\end{equation}
This central relation states that vortex stretching and vorticity diffusion tend to cancel for a fixed magnitude of vorticity on the statistical average. This balance has been extensively discussed by Novikov and co-workers \cite{novikov93jfr,novikov94mpl,mui96pre}. Recently its relation to the shape and evolution of the vorticity PDF was investigated \cite{wilczek09pre}.\par
The conditional balance is much more informative than the ordinary enstrophy balance as we, for example, can discuss the results as a function of vorticity magnitude highlighting possible correlations. Of course, the average enstrophy balance (discussed, e.g., by Tennekes and Lumley \cite{tennekes72book}) is obtained from equation \eqref{eq:balance} by multiplying with $\bs \Omega \, f(\bs \Omega)$ and integrating out $\bs \Omega$,
\begin{align}\label{eq:conditionaltoordinaryenstrophybalance}
  \big\langle \bs \omega \cdot \mat S \bs \omega \big\rangle + \big\langle \nu \bs \omega \cdot \Delta \bs \omega \big\rangle &= \int \bs \Omega \cdot \left[ \big\langle \mat{S}\bs \omega \big | \bs \Omega \big\rangle + \big\langle \nu \Delta \bs \omega \big | \bs \Omega \big\rangle \right]f(\bs \Omega) \, \mathrm{d}\bs \Omega \\
  &= \int_0^{\infty} \Omega\left[ s(\Omega) + d(\Omega) \right]\check f(\Omega) \, \mathrm{d}\Omega \approx 0 .
\end{align}
Higher-order moment relations can also be obtained in the same manner. The fact that conditional averaging allows to study the vorticity budget as a function of the vorticity eventually permits to discriminate the statistics of strong vorticity events from weak vorticity events as demonstrated later on.\par
If we now want to establish a connection between this conditional balance and the coherent vorticity structures present in the flow, we have to discriminate coherent from incoherent contributions to the vorticity field. The challenge in this context is to define what exactly a coherent structure is, and many different approaches have been introduced in recent years \cite{wallace90amr,dubief00jot}. A conceptually different way to determine the coherent contributions is to analyze the vorticity field in terms of CVE introduced by Farge and co-workers \cite{Farge_1999,Farge_2001, Farge_2001_2} which is based on a denoising approach.\par
To this end, we decompose the vorticity field into coherent and incoherent contributions according to
\begin{equation}
  \bs \omega(\bs x,t)=\bs \omega^c(\bs x,t) + \bs \omega^i(\bs x,t)
\end{equation}
using an orthogonal wavelet decomposition as described in section \ref{sec:cve}. By this decomposition we obtain for the conditional diffusive term
\begin{equation}
  \big\langle \nu \Delta \bs \omega \big | \bs \Omega \big\rangle = \big\langle \nu \Delta \bs \omega^c \big | \bs \Omega \big\rangle + \big\langle \nu \Delta \bs \omega^i \big | \bs \Omega \big\rangle = \left[ d^c(\Omega) + d^i(\Omega) \right] \widehat{\bs \Omega},
\end{equation}
such that we get two separate contributions from the coherent and incoherent parts of all fields of the ensemble. This decomposition, for instance, allows to quantify how much enstrophy dissipation is contained in the two terms.\par
The nonlinear vortex stretching term turns out to be more complicated as the rate-of-strain tensor contains both coherent and incoherent contributions. This can be seen by noting that we can calculate the coherent and incoherent velocity, $\bs u^c(\bs x,t)$ and $\bs u^i(\bs x,t)$, from the decomposed vorticity field via Biot-Savart's law and subsequently the coherent and incoherent rate-of-strain tensor $\mat{S}^c(\bs x,t)$ and $\mat{S}^i(\bs x,t)$. Hence, the vortex stretching term may be split up into four terms containing all possible combinations of coherent and incoherent parts of the vorticity and the rate-of-strain tensor,
\begin{align}
  \big\langle \mat{S}\bs \omega \big | \bs \Omega \big\rangle &= \big\langle \mat{S}^c\bs \omega^c \big | \bs \Omega \big\rangle + \big\langle \mat{S}^c\bs \omega^i \big | \bs \Omega \big\rangle + \big\langle \mat{S}^i\bs \omega^c \big | \bs \Omega \big\rangle + \big\langle \mat{S}^i\bs \omega^i \big | \bs \Omega \big\rangle \\
  &= \left[ s^{cc}(\Omega) + s^{ci}(\Omega) + s^{ic}(\Omega) + s^{ii}(\Omega) \right] \widehat{\bs \Omega}.
\end{align}
The interesting fact now is that the different functions characterize the interaction of coherent and incoherent contributions of the rate-of-strain field with coherent and incoherent contributions of the vorticity field. This will, for instance, eventually allow to quantify if vortex stretching is mainly caused by coherent structures or not.\par
It is also useful to compare the results obtained by wavelet filtering to another more classical filtering technique. To this end we make use of Fourier decomposition into ideal low pass filtered and high pass filtered contributions of the vorticity field according to
\begin{equation}
  \bs \omega(\bs x,t)=\bs \omega^l(\bs x,t) + \bs \omega^h(\bs x,t) .
\end{equation}
The cutoff wavenumber is determined such that the number of Fourier modes representing the low pass filtered vorticity field approximately corresponds to the number of wavelet coefficients representing the coherent vorticity field. More details on the wavelet filtering technique will be given in section \ref{sec:cve}.

\section{Coherent Vorticity Extraction}\label{sec:cve}
In the following, the notation for the orthogonal wavelet decomposition of a three-dimensional vector-valued field is presented, and the main ideas of CVE are explained. For more details, the reader is referred to the original paper by Farge et al. \cite{Farge_2001}.\par
The vector field vorticity is considered, $\bs{\omega}(\bs{x})=(\omega_1,\omega_2,\omega_3)$, $\bs{x}=(x_1,x_2,x_3)~~\in~~\lbrack 0,2\pi \rbrack^3$ given at resolution $N=2^{3J}$, where $J$ is the number of octaves in each space direction. The decomposition of $\bs{\omega}$ into an orthogonal wavelet series yields
\begin{eqnarray}
\bs{\omega}(\bs{x})= \sum_{\bs{\lambda} \epsilon \bs{\Lambda}} \widetilde{\bs{\omega}}_{\bs{\lambda}} \, {\psi}_{\bs{\lambda}}(\bs{x}),
\label{Eq: WAVE1}
\end{eqnarray}
where the multi-index $\bs{\lambda}=(j,i_x,i_y,i_z,\mu)$ denotes the scale index $j$, the position index $\bs{i}=(i_x,i_y,i_z)$, and the seven directions $\mu=1,2,...,7$ of the wavelets, respectively. The corresponding index set $\bs{\Lambda}$ is given by
\begin{eqnarray}
 \bs{\Lambda}=\lbrace \bs{\lambda}=(j,i_x,i_y,i_z,\mu);~j=0,...,J-1; ~i_x,i_y,i_z=0,...,2^J-1; \text{ and } \mu=1,...,7 \rbrace.
\end{eqnarray}
The orthogonality of the chosen wavelets implies that the coefficients are given by $\widetilde {\bs \omega}_{\bs \lambda} = \int_{[0, 2 \pi]^3}
{\bs \omega}({\bs x}) \psi_{\bs \lambda}({\bs x}) \, \mathrm{d} {\bs x}$. The coefficients measure fluctuations of $\bs{\omega}$ around scale $2^{-j}$ and around position $2 \pi \bs i/2^j$ in one of the seven possible directions $\mu$. The fast wavelet transform, which has linear complexity, is used to compute efficiently the $N-1$ wavelet coefficients $\widetilde{\bs{\omega}}_{\bs{\lambda}}$ (and the mean value which vanishes in the present case) from the $N$ grid point values of $\bs{\omega}$. The Coiflet 30 wavelet \cite{daubechies92cap} is chosen in this study; it has $10$ vanishing moments and is well adapted to our case. In past publications\cite{Farge_1999,Okamoto_2007} also Coiflet 12 wavelets have been used. It has, however, been checked that statistical results are robust with respect to this particular choice. The main steps of the CVE method and our analysis are summarized in the following:\\ 
\begin{itemize}
\item[i)] The wavelet coefficients of the vorticity field, $\widetilde{\bs{\omega}}_{\bs{\lambda}}$, are obtained by computing the fast wavelet transform of each component of the vorticity vector.
\item[ii)] Then we threshold the wavelet coefficients $\widetilde{\bs{\omega}}_{\bs{\lambda}}$, and the field can thus be split into two contributions. The wavelet coefficients of the coherent part are defined as
\begin{equation}
\widetilde{{\bs{\omega}}}_{\bs{\lambda}}^c = \left\lbrace
\begin{array}{cc}
\widetilde{{\bs{\omega}}}_{\bs{\lambda}} & ~~~~ {\rm if} ~~~~ \lVert\widetilde{{\bs{\omega}}}_{\bs{\lambda}}\rVert > \varepsilon= \sqrt{2 \sigma_f^2 \ln{N}} \\
\bs 0 & ~~~~ {\rm else}
\end{array}
\right.
\end{equation}
and those of the incoherent part as the remainder. Here $\sigma_f=\sqrt{\langle (\bs \omega^i)^2 \rangle_f/3}$ is the standard deviation of the individual incoherent vorticity field ($\langle \cdot \rangle_f$ denotes the spatial average over a single field) and $N$ the total number of grid points.
The threshold $\varepsilon$ is motivated as an optimal method for denoising signals of inhomogeneous regularity \cite{Donoho_1994} and is sometimes called 'universal' as it does not depend on the signal, but only on the sampling size and on the variance of the noise. However, as the variance of the noise, which equals two-third the enstrophy of the incoherent field, is unknown, a first threshold is calculated from the total field and the thresholding is applied. In the next step, the total field is then split with a threshold calculated from the incoherent field. An iterative algorithm was developed to obtain the optimal threshold by Azzalini et al. \cite{Azzalini_2005}. In the present work we have chosen to perform  one iteration in order to privilege a good compression rate rather than a perfectly denoised contribution.
\item[iii)] The fast inverse wavelet transform is applied to reconstruct the coherent vorticity $\bs{\omega}^c$. The incoherent vorticity $\bs{\omega}^i$ is obtained by subtracting $\bs \omega^c$ pointwise, such as $\bs{\omega}^i=\bs{\omega}-\bs{\omega}^c$. Since by construction the two fields $\bs{\omega}^c$ and $\bs{\omega}^i$ are orthogonal, the separation of the total enstrophy $Z=\langle \bs \omega^2 \rangle/2$ into $Z=Z^c+Z^i$ is ensured because the interaction term $\int_{\lbrack 0,2 \pi \rbrack^3} \bs{\omega}^c(\bs x)\cdot\bs{\omega}^i(\bs x) \, \mathrm{d}\bs{x}$ vanishes.
\item[iv)] Biot-Savart's law $\bs{u}=-\nabla \times (\nabla^{-2} \bs{\omega})$ is used to obtain the corresponding total, coherent and incoherent velocities, where again we have $\bs{u}=\bs{u}^c + \bs{u}^i$. However, as the Biot-Savart operator is not diagonal in wavelet space, the decomposition of the turbulent kinetic energy $E=\langle \bs{u}^2 \rangle/2$ is $E=E^c+E^i+\epsilon_{E}$ where $\epsilon_{E}\ne 0$ but small \cite{Farge_1999}.\\
\item[v)] From the decomposed velocity field we then can calculate the total rate-of-strain tensor $\mat S=\frac{1}{2}\left[ \nabla\bs u +(\nabla \bs u)^T \right]$, as well as the coherent and incoherent contributions according to $\mat S^c=\frac{1}{2}\left[ \nabla\bs u^c +(\nabla \bs u^c)^T \right]$ and $\mat S^i=\frac{1}{2}\left[ \nabla\bs u^i+(\nabla \bs u^i)^T \right]$, respectively.\\
\item[vi)] Finally, also the Laplacian of total, as well as the coherent and incoherent contributions of the vorticity field is calculated.
\end{itemize}

\section{Direct Numerical Simulation}

\begin{table}
\begin{center}
  \begin{tabular}{cccccccc}
    \hline 
    $N$ & $R_{\lambda}$ & $u_{\mathrm{rms}}$ & $\nu$ & $L$ & $T$ & $\eta$ & $\tau_{\eta}$\\
    \hline 
    $512^3$ \hspace{.1cm}& $112$ \hspace{.1cm}& $0.543$ \hspace{.1cm}& $0.001$ \hspace{.1cm}& $1.55$ \hspace{.1cm}& $2.86$ \hspace{.1cm}& $9.92 \cdot 10^{-3}$ \hspace{.1cm}& $9.85 \cdot 10^{-2}$\\
    \hline
  \end{tabular}
\end{center}
\caption{Simulation parameters. Number of grid points $N$, Reynolds number based on the Taylor micro-scale $R_{\lambda}$,
  root-mean-square velocity $u_{\mathrm{rms}}$, kinematic viscosity $\nu$, integral length scale $L$, large-eddy turnover time $T$, Kolmogorov length scale $\eta$, and Kolmogorov time scale $\tau_{\eta}$.}
\label{tab:simpara}
\end{table}
The presented flow is generated with a standard, dealiased Fourier pseudospectral code \cite{canuto87book,hou2007jcp} for the vorticity equation. The integration domain is a triply periodic box of side-length $2\pi$. The time stepping scheme is a third-order Runge-Kutta scheme \cite{shu88jcp}. For the statistically stationary simulations a large-scale forcing is applied to the flow. Here, care has to be taken in order to fulfill the statistical symmetries we make use of in our theoretical framework \cite{wilczek11jfm}. After numerous tests, we chose a large-scale forcing which conserves the kinetic energy of the flow by amplifying the magnitude of Fourier modes in a wavenumber band and letting their phases evolve freely. This forcing has been found to  deliver satisfactory results concerning the statistical symmetries. The numerical results presented in the following are obtained from a simulation whose simulation parameters are listed in table \ref{tab:simpara}.\par
For the numerical evaluation we average over twenty realizations of the vorticity field to ensure a good statistical quality. Furthermore, the presented data stems from a well-resolved simulation ($k_{\mathrm{max}}\eta \approx 2$), which is necessary as, e.g., second derivatives of the vorticity field will be considered. We also refer to the discussion in the literature \cite{Yakhot_2005,Schumacher_2007} regarding the importance of the spatial resolution.

\section{Results}

\begin{figure}
  \includegraphics[width=1.0\textwidth]{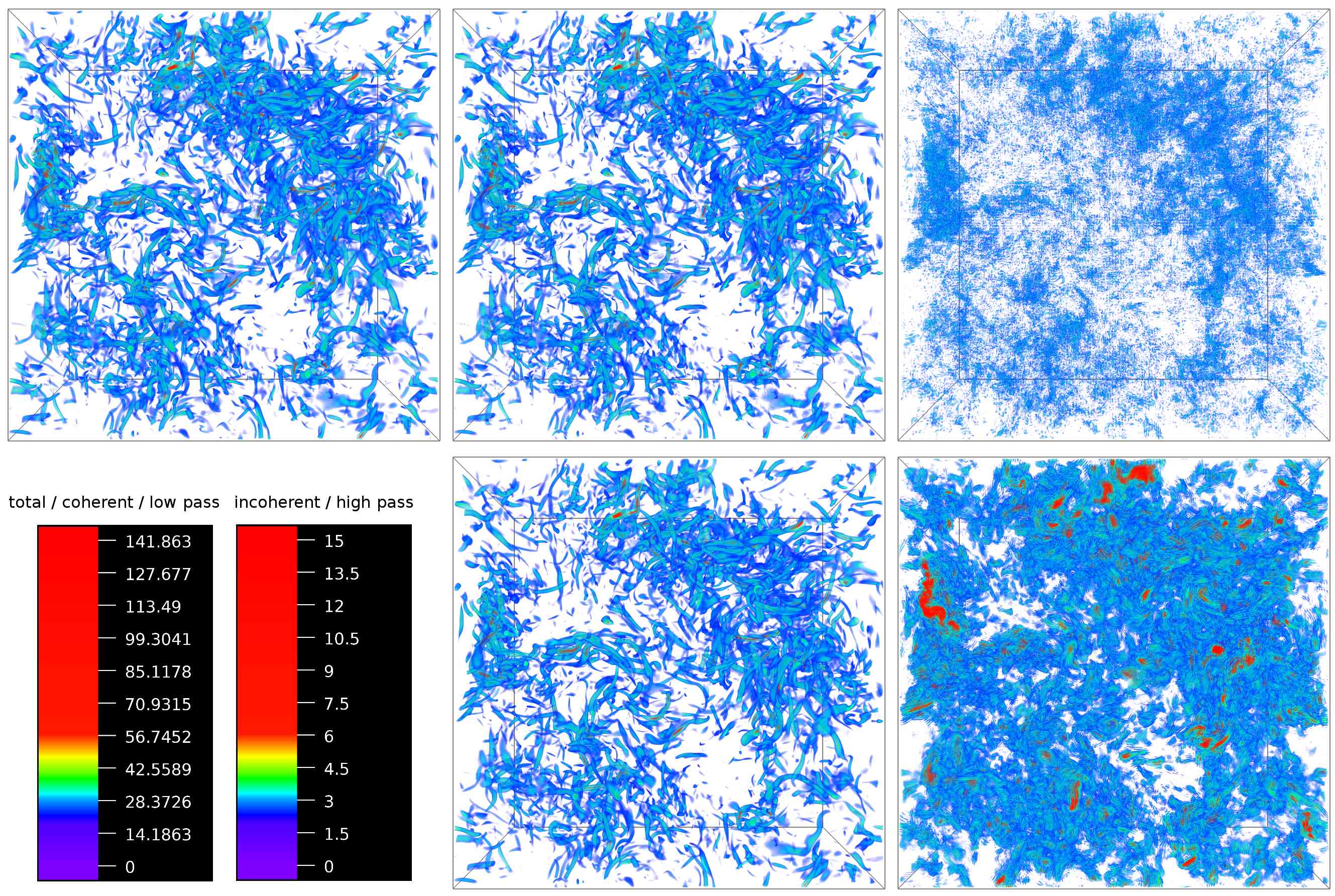}
  \caption{Volume visualization of a wavelet-decomposed (upper panels) and Fourier-decomposed (lower panels) vorticity field (shown are the magnitudes). The upper left shows the total field, the pictures in the middle show the coherent/low pass filtered contribution, whereas the right pictures show the incoherent/high pass filtered contributions. The lower left picture indicates the different color scales used (the standard deviation of the ensemble of total fields is $\sigma=\sqrt{\langle \bs \omega^2 \rangle/3}=5.87$). The visualizations have been produced with the free software package VAPOR (\texttt{www.vapor.ucar.edu}).}\label{fig:viz}
\end{figure}

\begin{figure}
  \includegraphics[width=1.0\textwidth]{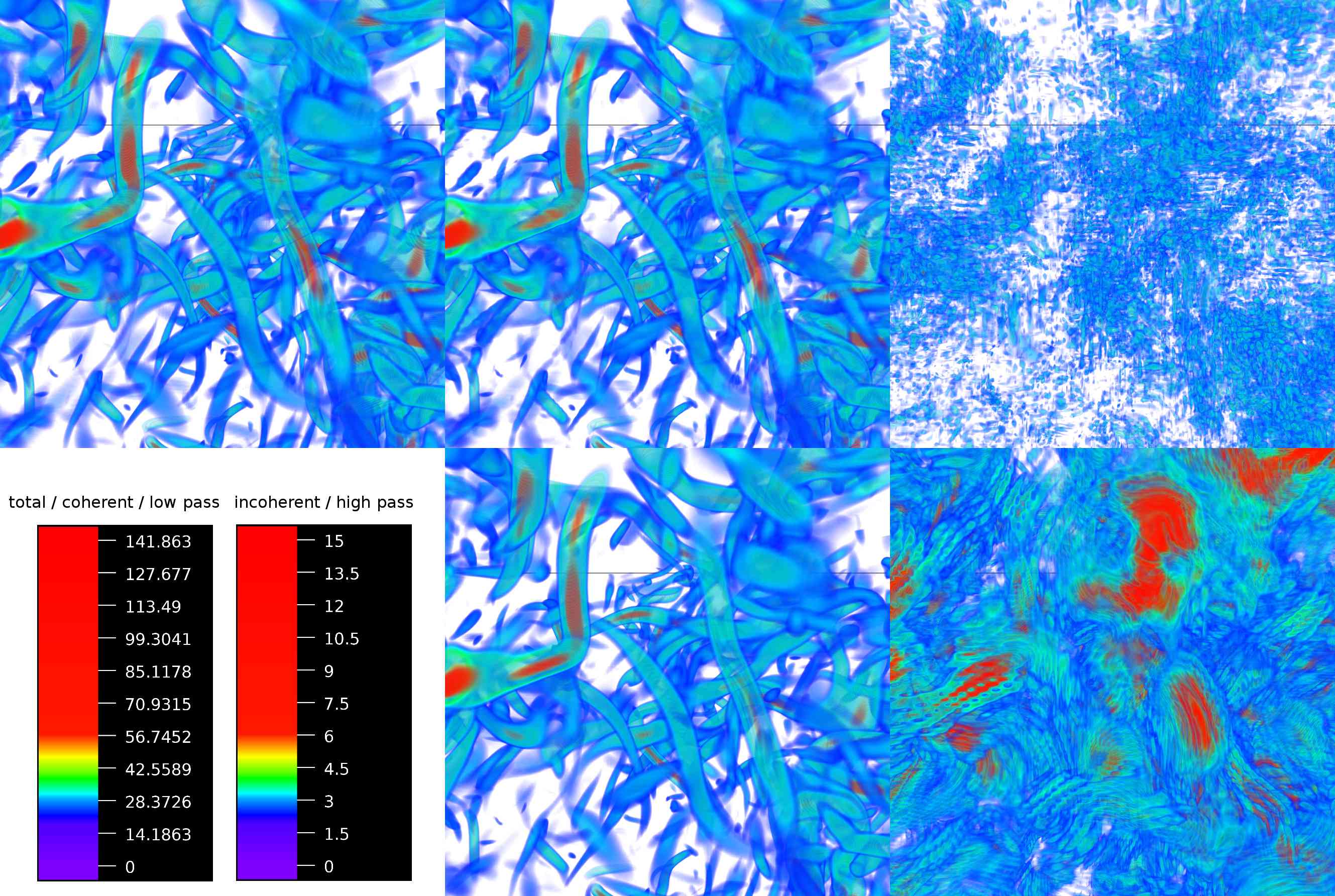}
  \caption{Same visualization as in figure \ref{fig:viz}, but in a close-up view. Both the coherent and low pass filtered contributions represent the structure of the vorticity field rather well, the low pass filtered contribution tends to underestimate, for example, the core of coherent vortex structures. These contributions are contained in the high pass filtered contribution.}\label{fig:viz2}
\end{figure}

\begin{figure}
  \includegraphics[width=.49\textwidth]{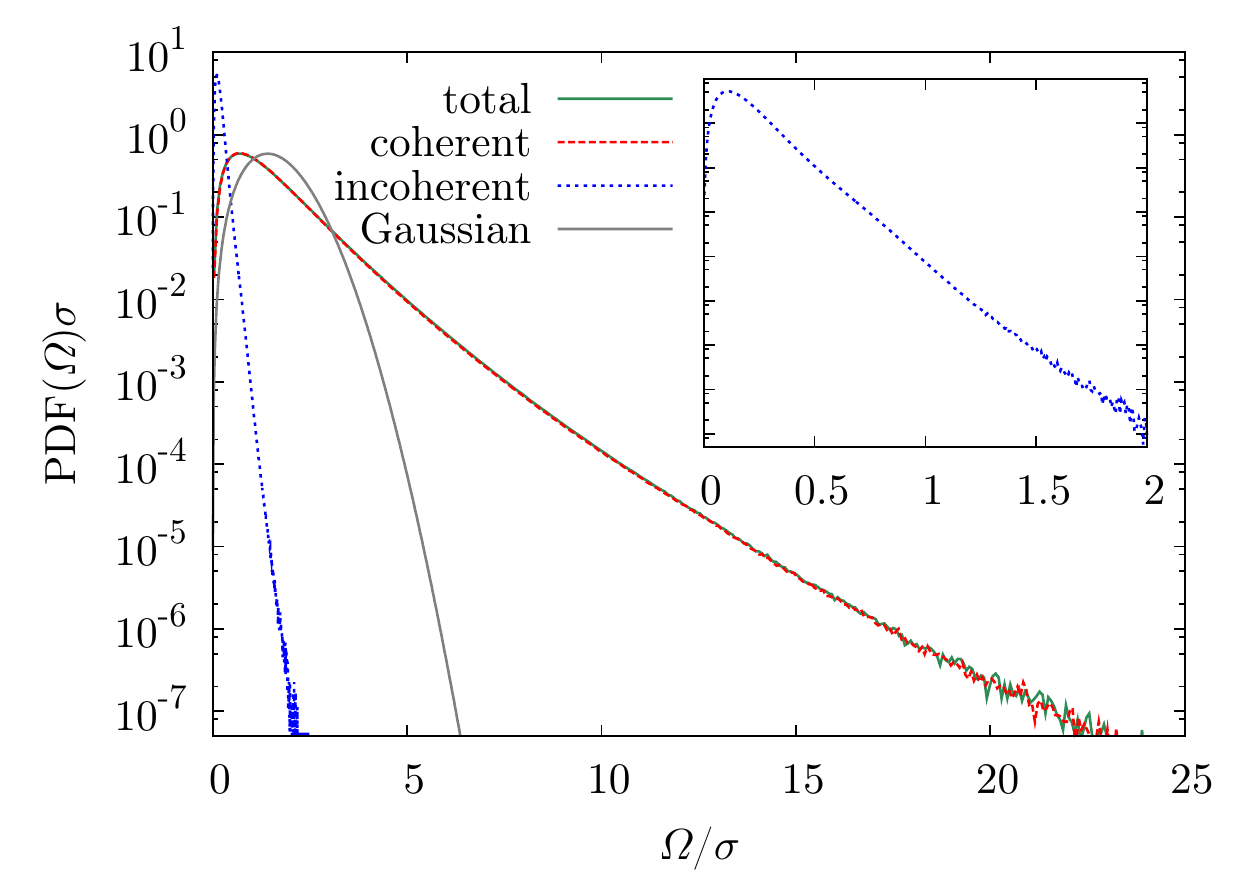}
  \includegraphics[width=.49\textwidth]{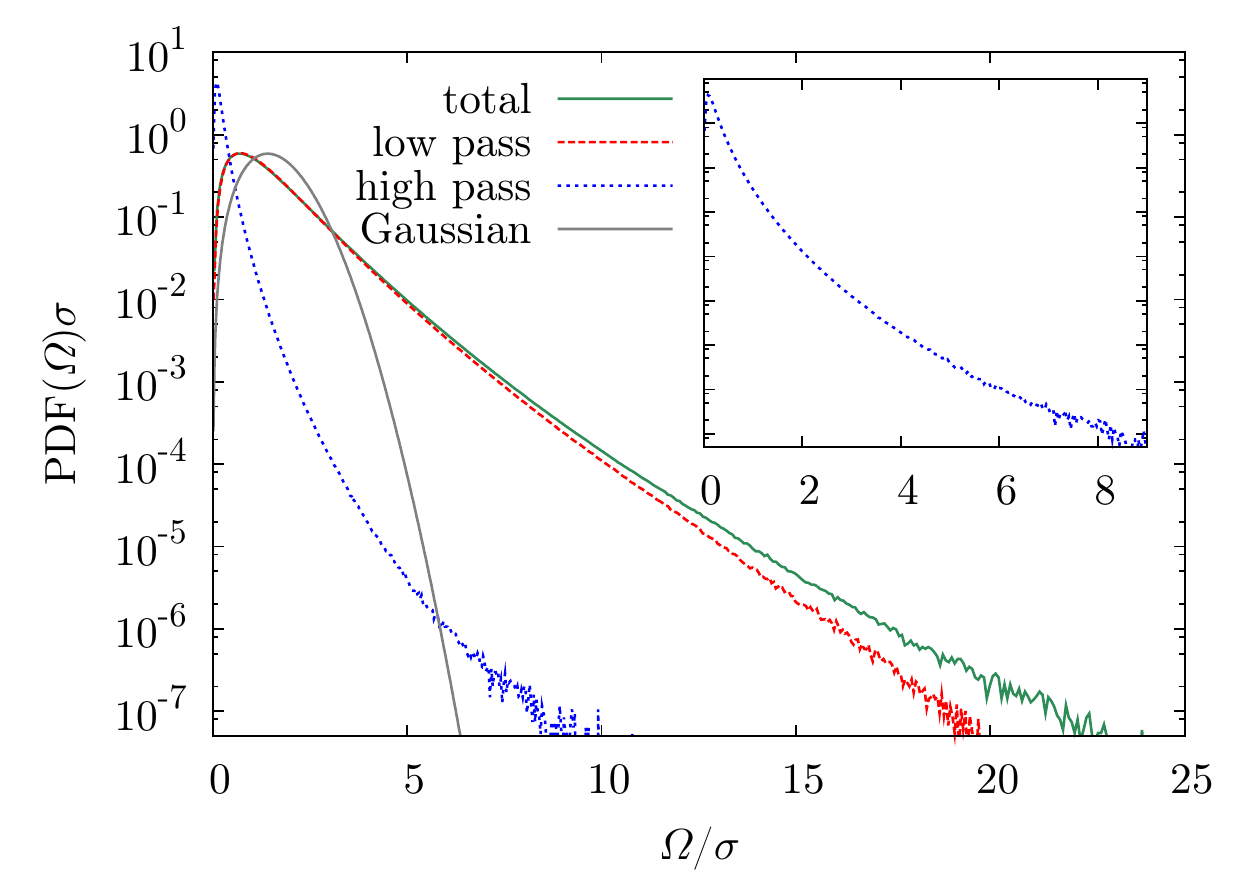}
  \caption{Left: PDFs of the magnitude of the total vorticity, the coherent part and the incoherent part of the wavelet-decomposed fields. Right: PDFs of the total, low pass filtered and high pass filtered Fourier-decomposed fields. The PDFs are highly non-Gaussian with stretched exponential tails in the case of the total and coherent vorticity. The incoherent part displays a much lower amplitude. Inset: Close-up of the PDF of the incoherent/high pass filtered vorticity. The nearly exponential/stretched exponential tail of the PDF indicates that also the incoherent/high pass filtered part is non-Gaussian. Note that the PDFs have been normalized by the standard deviation of the total vorticity $\sigma$. The magnitude PDF corresponding to a Gaussian distributed vorticity field with standard deviation $\sigma$ is shown for reference.}\label{fig:pdfs}
\end{figure}

\begin{figure}
  \includegraphics[width=.49\textwidth]{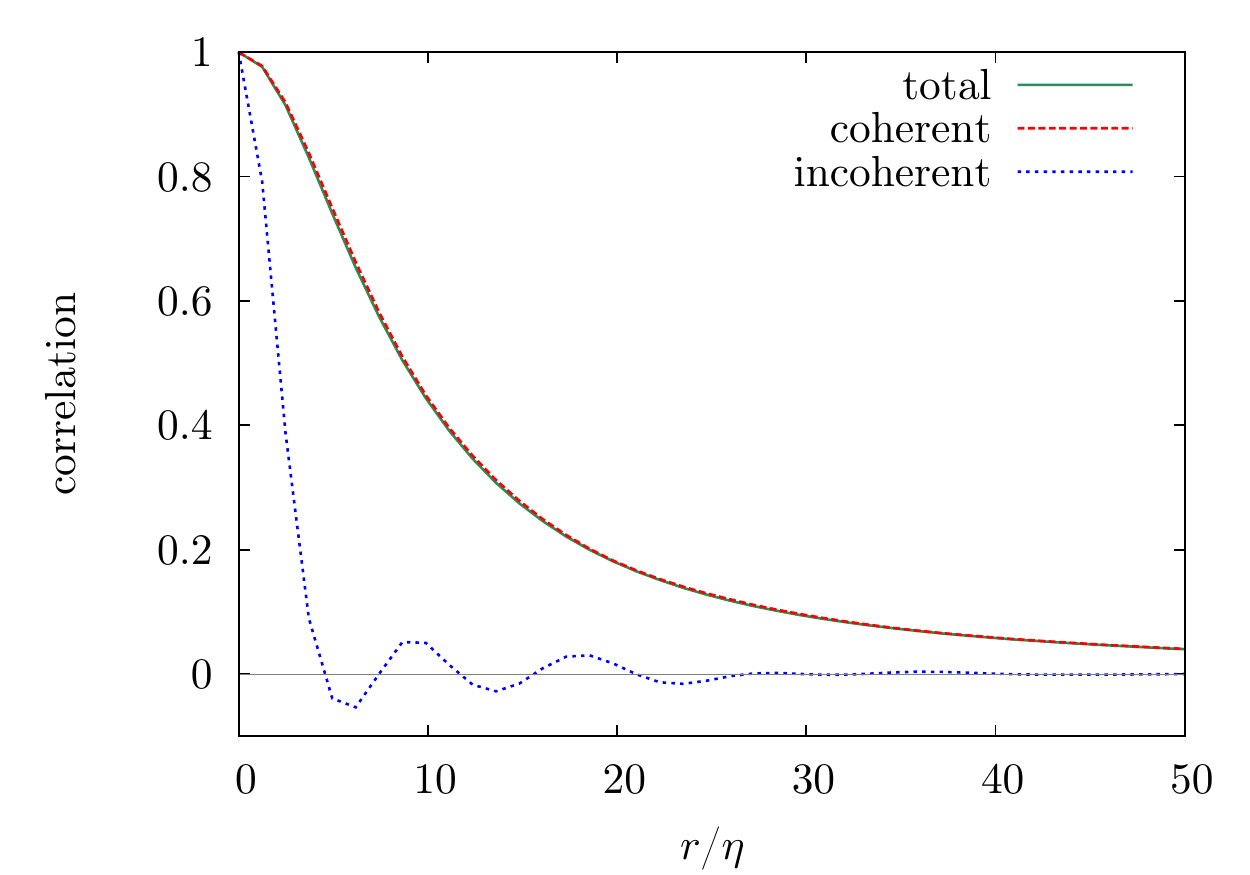}
  \includegraphics[width=.49\textwidth]{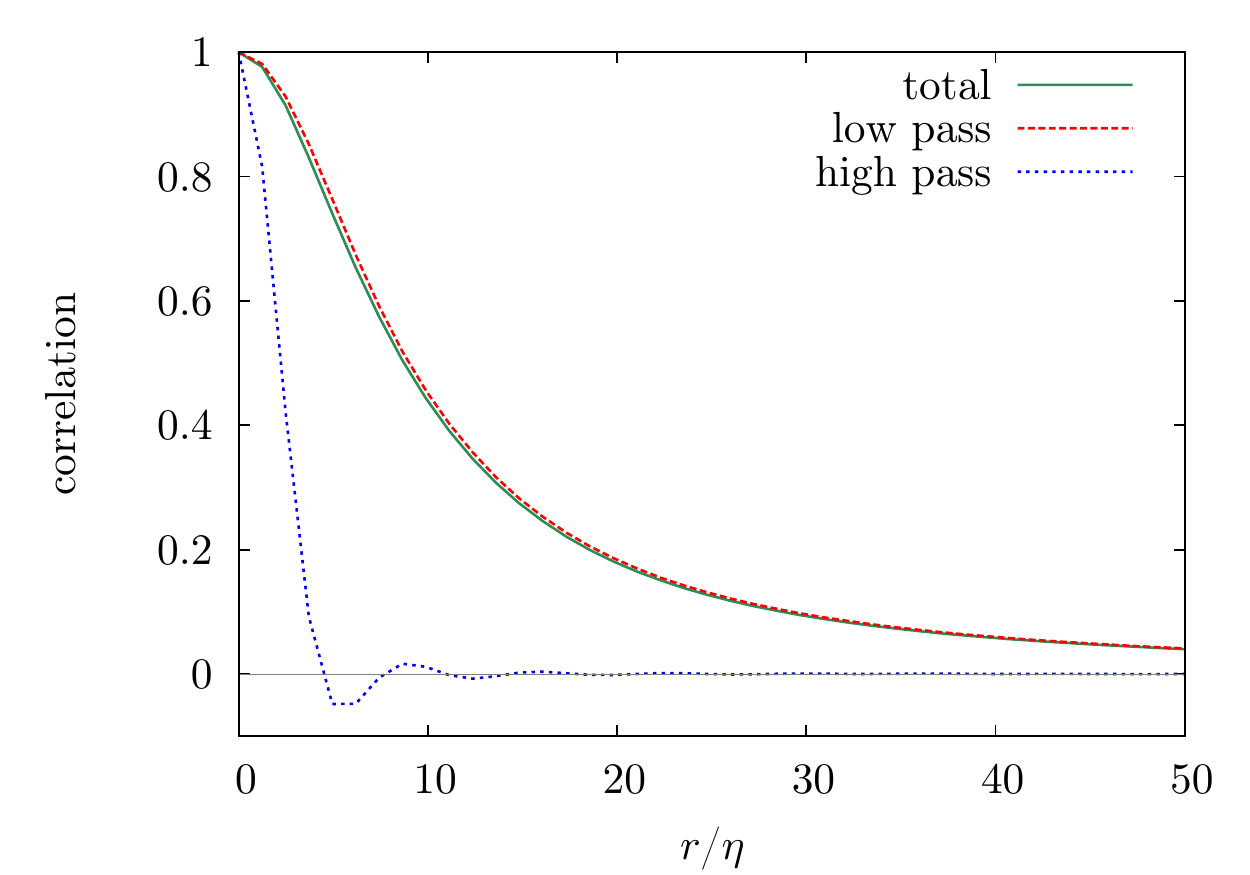}
  \caption{Longitudinal vorticity autocorrelation functions $\langle \omega_x(\bs x) \, \omega_x(\bs x+r\bs e_x) \rangle/\langle \omega_x(\bs x)^2 \rangle$ for the total, coherent/low pass filtered and incoherent/high pass filtered vorticity (left: wavelet decomposition, right: Fourier decomposition). While the total and coherent/low pass correlations almost coincide, the autocorrelation function of the incoherent/high pass filtered part is rapidly decaying and oscillating. The incoherent part of the wavelet-decomposed fields is longer correlated than the high pass filtered part of the Fourier-decomposed fields. It can be seen that the incoherent/high pass filtered fields are much shorter correlated than the coherent/low pass filtered contributions.}\label{fig:correlations}
\end{figure}

\begin{figure}
  \includegraphics[width=.6\textwidth]{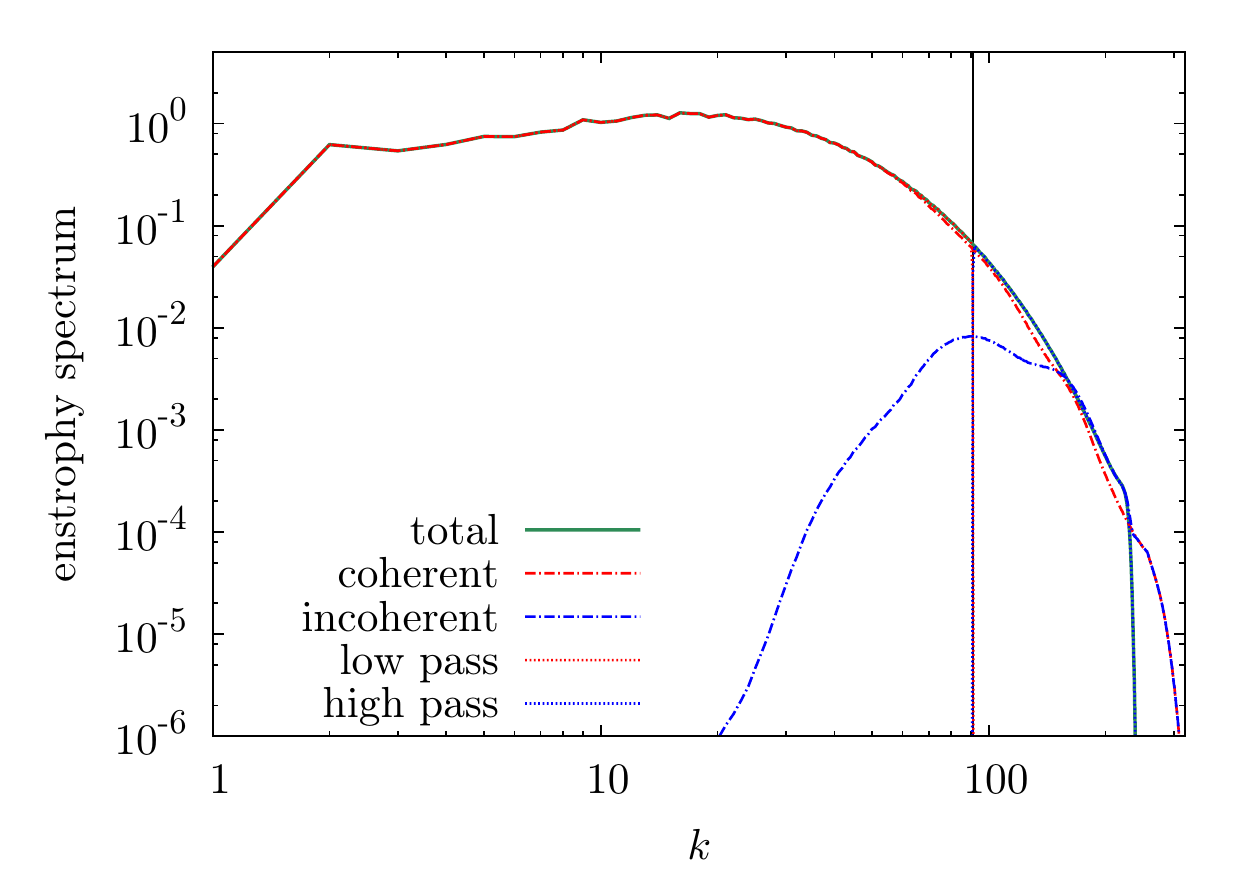}
  \caption{Enstrophy spectra for the total, coherent and incoherent contribution as well as the large-scale and small-scale contributions. The coherent spectrum matches the total one over a broad range of scales, small-scale deviations, however, are visible. The Fourier filter (vertical line at $k_c = 91$) clearly separates large-scale and small-scale contributions. The high-frequency contributions of the wavelet spectra in the coherent and incoherent flow are due to a change of the basis functions, i.e. from trigonometric polynomials (Fourier basis) to Coiflet 30 (wavelet basis), which are not compactly supported in spectral space.}\label{fig:spectra}
\end{figure}

\begin{figure}
  \includegraphics[width=.6\textwidth]{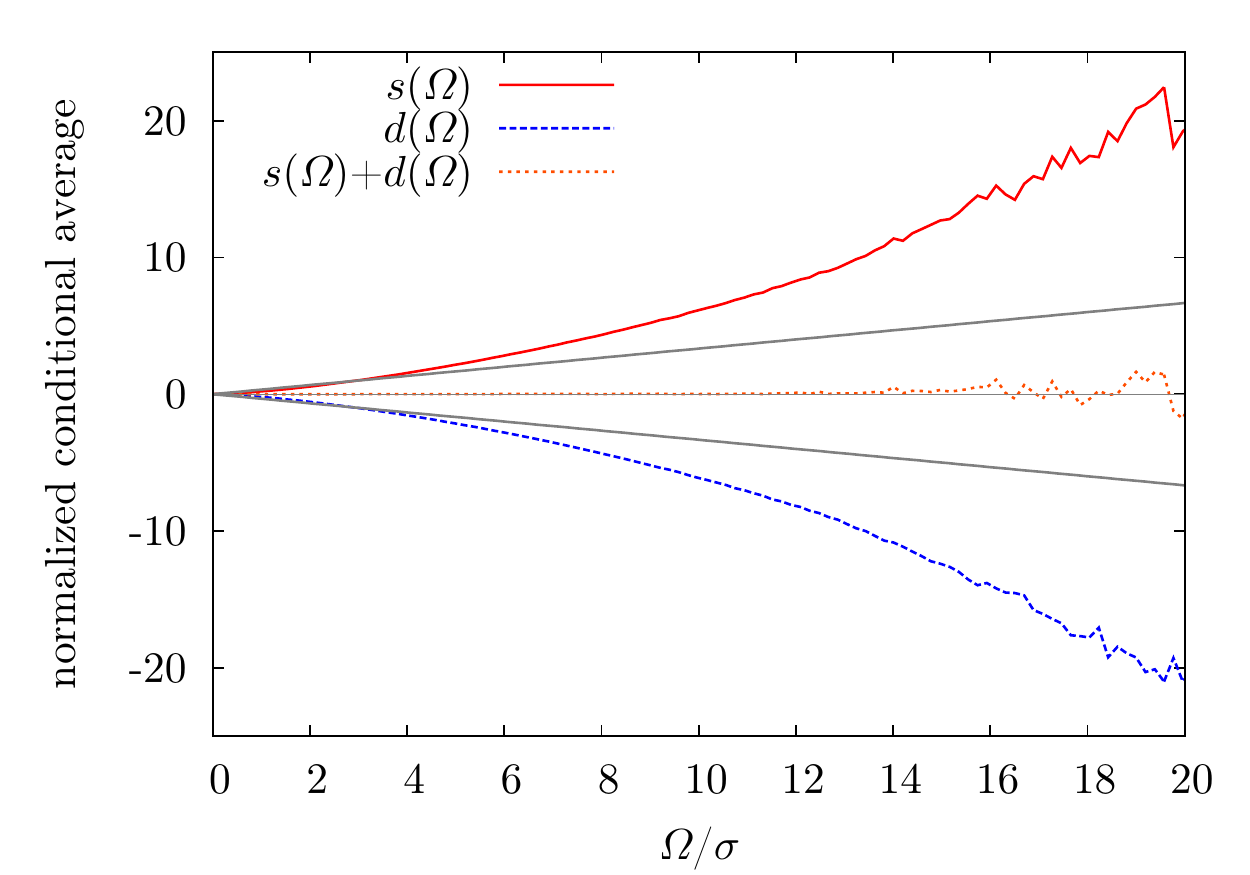}
  \caption{Conditional balance of the conditional averages related to vortex stretching and diffusion of vorticity. Vortex stretching is positively correlated with vorticity, whereas the diffusion of vorticity is negatively correlated. As expected, the sum of both terms cancels, indicating that the conditional balance \eqref{eq:balance} holds. The gray lines indicate the functional form of the conditional averages expected when assuming statistical independence of the rate-of-strain tensor and the vorticity magnitude and a corresponding balance of the diffusive term. All conditional averages are normalized by $\langle \varepsilon_{\omega} \rangle/\sigma$ consisting of the enstrophy dissipation and the standard deviation of the vorticity field, respectively.}\label{fig:balance}
\end{figure}

\begin{figure}
  \includegraphics[width=.49\textwidth]{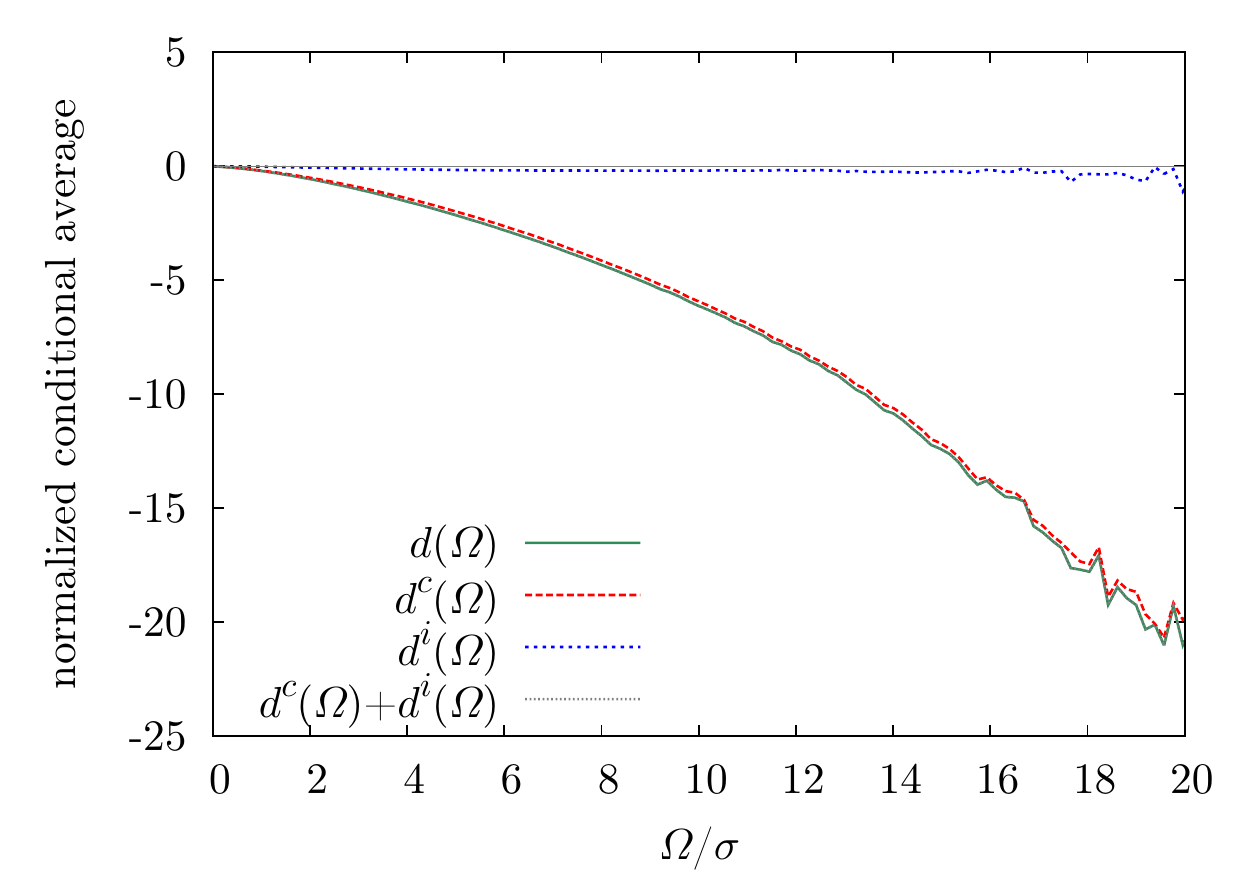}
  \includegraphics[width=.49\textwidth]{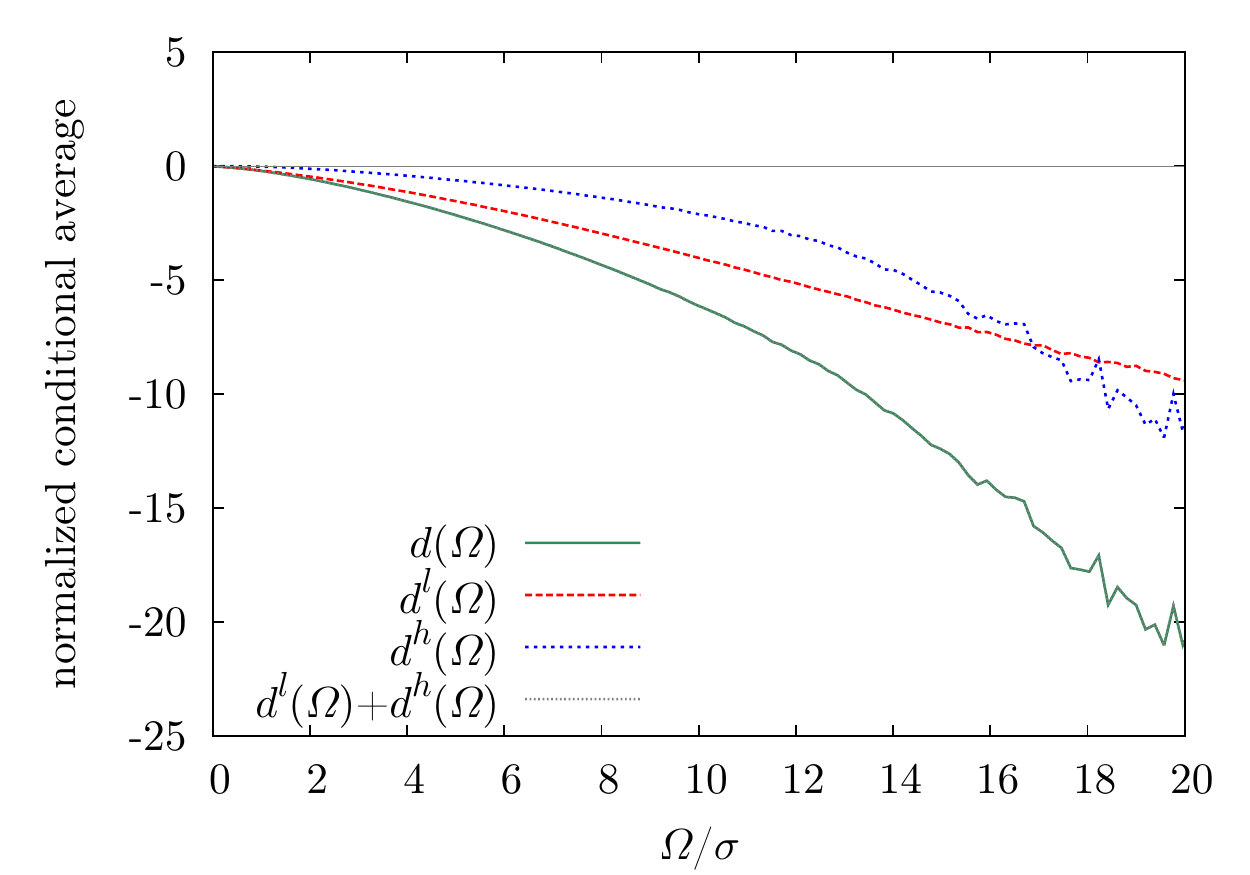}
  \caption{Coherent and incoherent parts of the diffusive term compared to the total one (left: wavelet decomposition, right: Fourier decomposition). In the case of the wavelet decomposition the coherent part contributes the most, however, the incoherent part is small, but non-vanishing. In the case of the Fourier-decomposed fields the low pass filtered and high pass filtered contributions do not separate that well. As a benchmark, the sum of both contributions is shown to add up to the total diffusive term, which has been calculated for reference.}\label{fig:diffusive}
\end{figure}

\begin{figure}
  \includegraphics[width=.49\textwidth]{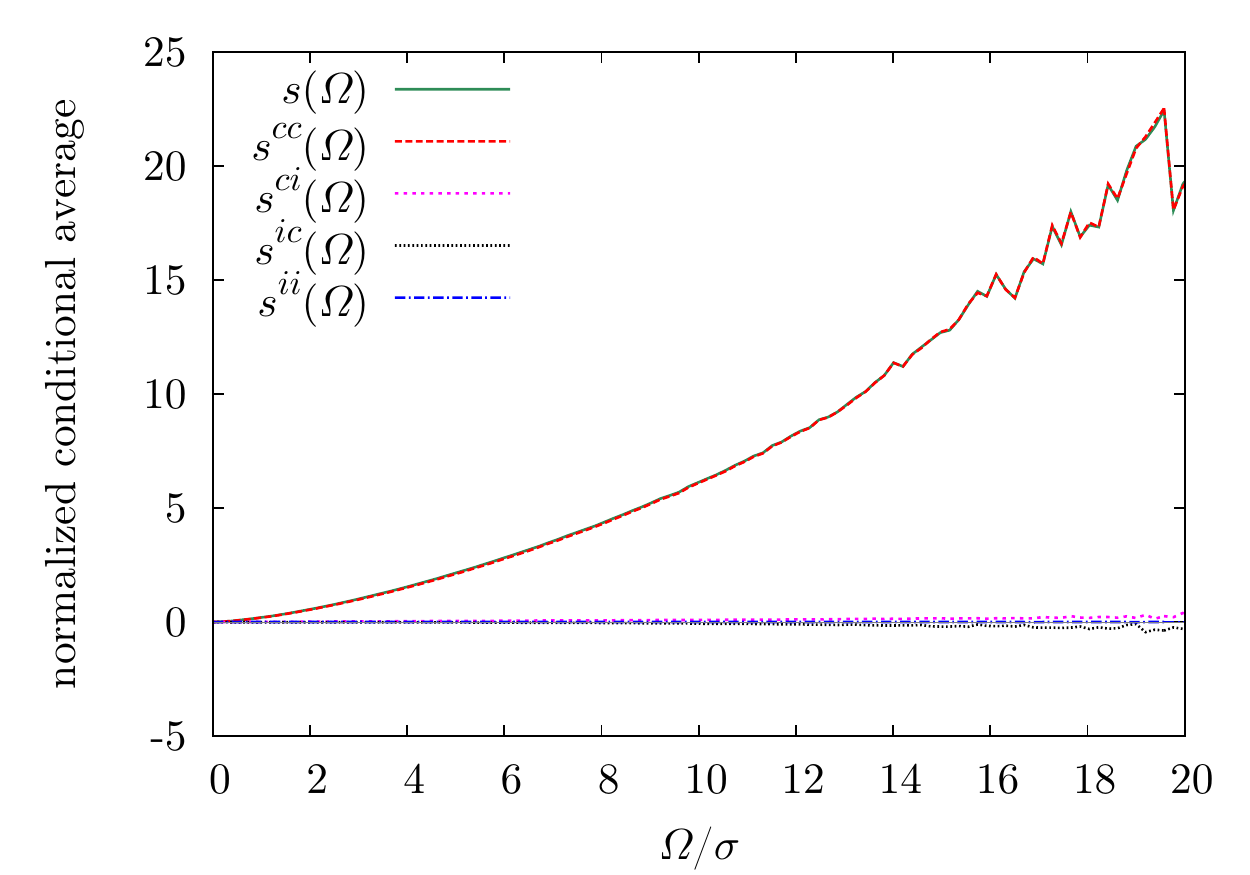}
  \includegraphics[width=.49\textwidth]{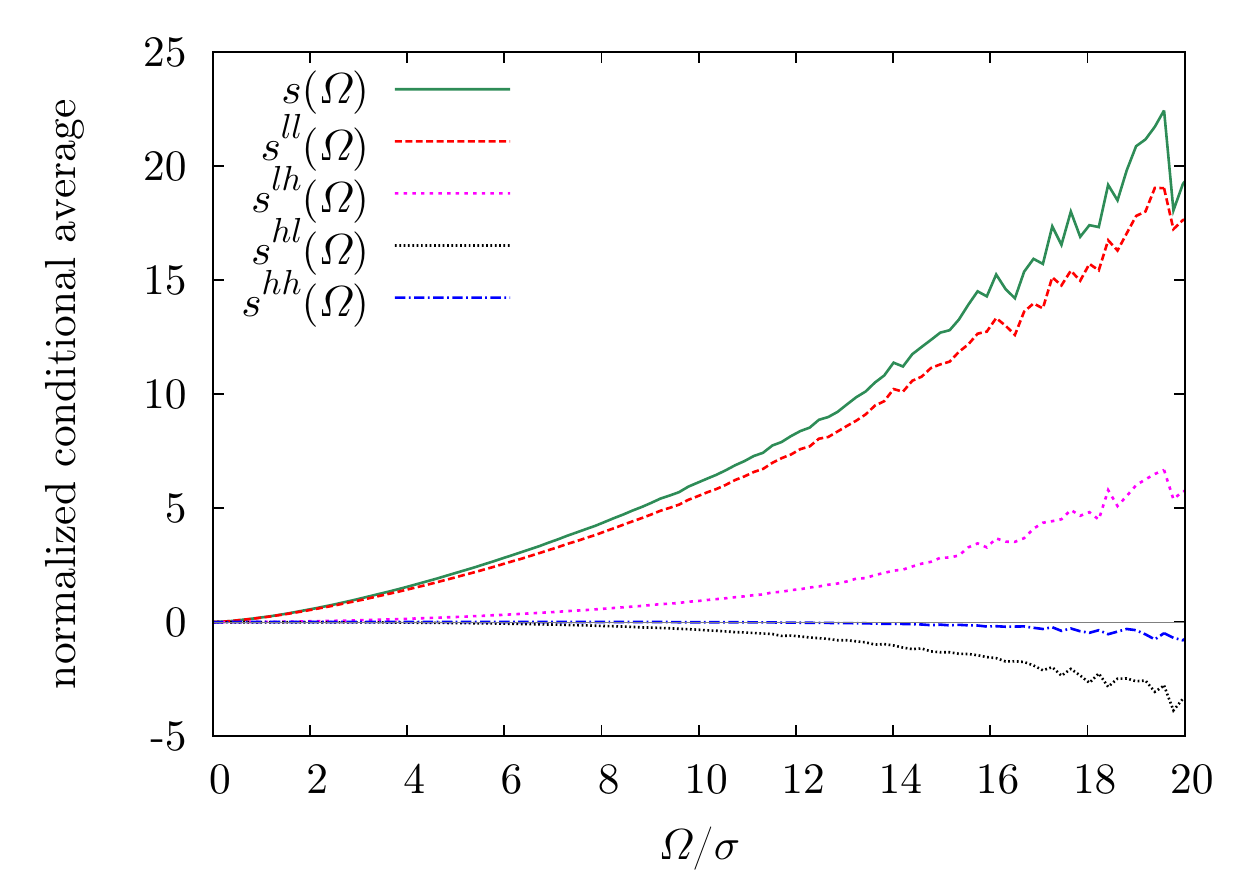}
  \caption{Coherent and incoherent contributions to the conditional average related to the vortex stretching term (left: wavelet decomposition, right: Fourier decomposition). For the wavelet-decomposed fields the coherent-coherent contribution is dominant and almost identical to the total term. This indicates that vortex stretching is predominantly caused by both the coherent vorticity and the rate-of-strain tensor induced by the coherent vorticity. For the Fourier-decomposed fields the low pass-low pass contribution deviates significantly from the total one.}\label{fig:vortexstretching}
\end{figure}

\begin{figure}
  \includegraphics[width=.49\textwidth]{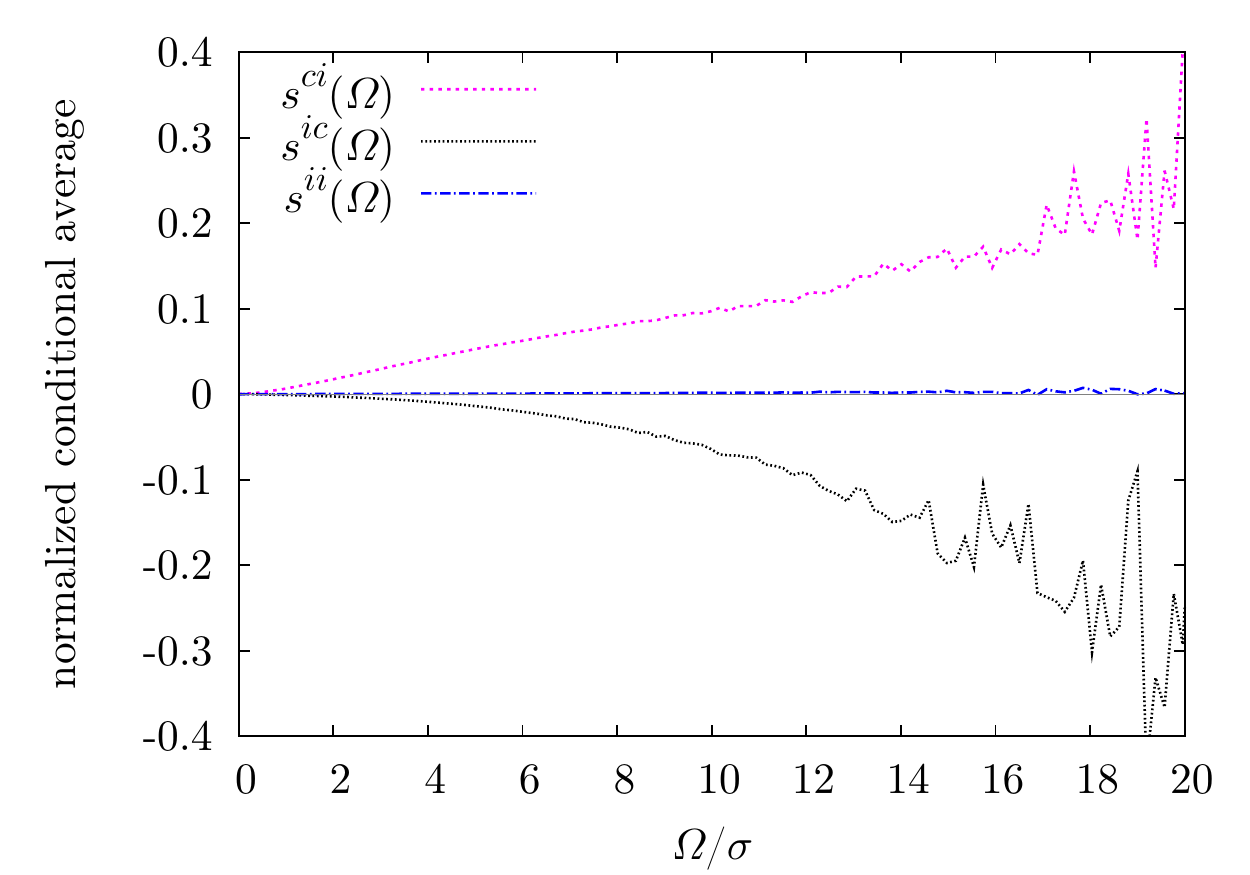}
  \includegraphics[width=.49\textwidth]{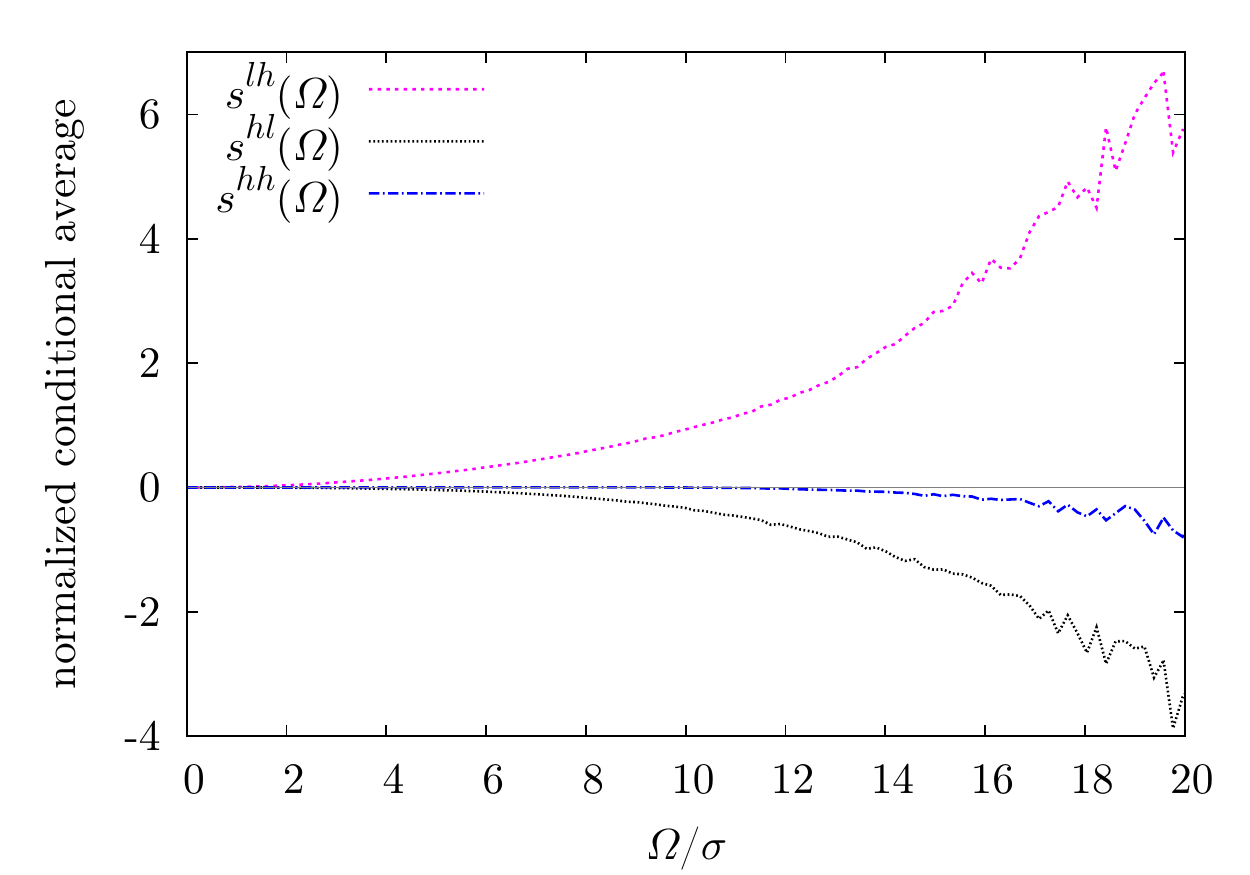}
  \caption{Contributions to the conditional vortex stretching term with at least one incoherent quantity (left: wavelet decomposition, right: Fourier decomposition). The conditionally averaged coherent part of the rate-of-strain tensor times the incoherent part of the vorticity contributes positively to the conditionally averaged vortex stretching term. In contrast to that, the term involving the incoherent rate-of-strain tensor times the coherent vorticity tends to deplete the vorticity. The term with both quantities incoherent seems to be negligible compared to the remaining terms. Note that the amplitude is significantly lower in the case of the wavelet-decomposed fields.}\label{fig:vortexstretchingincoherent}
\end{figure}

\begin{table}
\begin{center}
 \begin{tabular}{rccccc}
          \hline
           & $Z$   & $\%Z$ & min $\omega_{\ell}$  & max $\omega_{\ell}$ & \% of retained coeff.\\
           \hline
total:      & 51.6  & 100   & -118  & 123  & 100\\
           \hline
coherent:   & 51.0  & 98.8 & -118  & 122  & 2.31\\
           \hline
incoherent: & 0.643 &  1.25 & -10.7 & 10.5 & 97.7\\
      \hline
 \end{tabular}
\end{center}
\caption{Enstrophy and percentage of retained coefficients of the total, coherent and incoherent contributions obtained using CVE. Mean value of the threshold $\varepsilon=9.74$. $\omega_{\ell}$ denotes the components of vorticity. All values are averaged over the fields of the ensemble.}
\label{tab:cvestat}

\begin{center}
 \begin{tabular}{rccccc}
                      \hline
                    & $Z$   & $\%Z$ & min $\omega_{\ell}$   & max $\omega_{\ell}$ & \% of retained coeff.\\
                     \hline
total:               & 51.6  & 100   & -118  & 123  & 100\\
                    \hline
low pass filtered:   & 50.1  & 97.2 & -96.9 & 98.1 & 2.41\\
                    \hline
high pass filtered: & 1.47  &  2.85 & -47.7 & 49.5 & 97.6\\
      \hline
 \end{tabular}
\end{center}
\caption{Enstrophy and percentage of retained coefficients of the total, low pass and high pass filtered contributions obtained using Fourier filtering with cutoff wavenumber $k_c=91$, which was calculated as $k_c = \left( \frac{3}{4\pi} N \cdot c_r \right)^{1/3} \approx 91$, where $c_r=2.31 \%$ is the percentage of retained coefficients for the CVE. All values are averaged over the fields of the ensemble.}
\label{tab:fourstat}
\end{table}

\begin{table}
\begin{center}

  \begin{tabular}{rcccccc}
    \hline 
     & $\langle \bs \omega \cdot \mathrm{S}^{c/l} \bs \omega^{c/l} \rangle$ & $\langle \bs \omega \cdot \mathrm{S}^{c/l} \bs \omega^{i/h} \rangle$  & $\langle \bs \omega \cdot \mathrm{S}^{i/h} \bs \omega^{c/l} \rangle$ & $\langle \bs \omega \cdot \mathrm{S}^{i/h} \bs \omega^{i/h} \rangle$ & $\langle \nu \bs \omega \cdot \Delta \bs \omega^{c/l} \rangle$ & $\langle \nu \bs \omega \cdot \Delta \bs \omega^{i/h} \rangle$\\
    \hline 
    wavelet: & $0.978$ & $2.75 \cdot 10^{-2}$ & $-5.85 \cdot 10^{-3}$ & $3.71 \cdot 10^{-4}$ & $0.913$ & $8.70 \cdot 10^{-2}$\\
    \hline
    Fourier: & $0.931$ & $8.35 \cdot 10^{-2}$ & $-1.48 \cdot 10^{-2}$ & $4.62 \cdot 10^{-4}$ & $0.764$ & $0.236$ \\
    \hline
  \end{tabular}
\caption{Relative contributions of the different decomposed terms of the budget equation \eqref{eq:balance} to the average enstrophy budget.}
\label{tab:enstrophybudget}
\end{center}
\end{table}

Before coming to the investigation of the conditional vorticity budget, we show some benchmark results. Details on the CVE and Fourier filtering are summed up in tables \ref{tab:cvestat} and \ref{tab:fourstat}. For example, it can be seen that with about $2.3\%$ retained coefficients still about $99 \%$ of the enstrophy is contained in the coherent contribution. Due to the high cutoff wavenumber the Fourier filtering performs comparable. Furthermore, CVE represents the minimum and maximum values of the vorticity components to a better extent than the Fourier filtering. In figure \ref{fig:viz} volume visualizations of the total, coherent, incoherent, low pass filtered and high pass filtered contributions of the vorticity are shown. It can be seen that both the coherent and low pass filtered contributions represent the global structure of the total vorticity field to a good extent, differences are only visible in the details. When comparing incoherent and high pass filtered contributions, differences become more pronounced. While the incoherent contributions appear very noisy and small in amplitude, the high pass filtered contributions have a much larger amplitude and appear more structured. The color scales for the visualization of the incoherent and high pass filtered fields have been adjusted to account for that issue. A zoom into the fields, however, also indicates differences in the low pass filtered contribution compared to the total field, as presented in  figure \ref{fig:viz2}. Although the shape of the vortex structures is captured quite well, it can, for example, be seen that the amplitude of the vortex cores sometimes is underestimated in the low pass filtered contribution.\par
This observation can be made more quantitative by investigating the PDF of the magnitude of the vorticity which is presented in figure \ref{fig:pdfs}. This figure shows that the PDF of the coherent part of the vorticity yields a PDF almost indistinguishable from the total PDF. The PDF of the incoherent part has a largely reduced variance and displays a nearly exponential decay consistent with previous findings \cite{Farge_1999,Farge_2001,Farge_2003}. Compared to that, the Fourier filter does not perform as well because the PDF of the low pass filtered part of the vorticity field is clearly distinguishable from the total PDF \cite{Farge_2003}. Furthermore, the high pass filtered component of the vorticity field has a much larger variance compared to the PDF of the incoherent part and displays a stretched exponential shape.\par
To quantify the two-point statistics of the field, figure \ref{fig:correlations} shows the longitudinal correlation function of the different contributions. It can be seen that both the coherent and low pass filtered contributions of the vorticity field almost coincide with the correlation function of the total field. Differences, however, become apparent when comparing the incoherent and high pass filtered contributions. Although both correlation functions decay rapidly (compared to the correlation function of the total field), a long-ranging oscillation of the autocorrelation function of the incoherent field can be observed. These oscillations may also be seen in the visualizations (figures \ref{fig:viz} and \ref{fig:viz2}) in form of a very fine-scaled structure of the incoherent field.\par
These observations are also supported by studying the enstrophy spectra of the different contributions, c.f. figure \ref{fig:spectra}. The enstrophy spectra of the total and coherent flows perfectly superimpose all along the inertial range. In the dissipative range, for wavenumbers larger than $k>60$, we observe a departure, i.e., the enstrophy spectrum of the coherent flows decays faster than the one of the total flow. In contrary, the enstrophy spectrum of the incoherent flow has a much weaker amplitude and exhibits a slope close to $k^4$. After reaching its maximum value at $k=90$, it rapidly decays. For reason of comparison, we also plotted the cutoff wavenumber $k_c=91$ corresponding to the black vertical line; this line divides the enstrophy spectrum of the total flow into large-scale contributions $k<k_c$ and small-scale contributions $k\ge k_c$. According to the Wiener-Khinchin theorem, the spectra correspond to the Fourier transform of the autocorrelation functions. Hence, the oscillations observed in the longitudinal vorticity autocorrelation functions (see fig \ref{fig:correlations}) for both, the incoherent and the small-scale contribution are related to the maximum values in the corresponding spectra, i.e., the wavenumber of the oscillations is given by $k_{\mathrm{max}}=(2\pi)/l_r$ where $l_r$ denotes the wavelength of the oscillations and $k_{\mathrm{max}}$ the maximum in the spectrum.\par
We now come to the conditional vorticity budget and start with an investigation of the functional form of $s(\Omega)$ and $d(\Omega)$ presented in figure \ref{fig:balance}. It can be seen that the vortex stretching term is positively correlated with the vorticity, whereas the diffusive term is negatively correlated. This is physically quite intuitive as it mirrors the fact that the vortex stretching term tends to amplify vorticity, while the dissipative term depletes vorticity. The fact that the sum of both averages nearly identically vanishes represents \textit{a posteriori} justification for the approximation leading to the relation \eqref{eq:balance}. In the same figure the functions expected for the case where the rate-of-strain tensor is assumed statistically independent of the vorticity and the corresponding diffusive term balances this term are shown for comparison. The slope of these linear functions is obtained such that these functions yield the correct ordinary enstrophy budget \eqref{eq:conditionaltoordinaryenstrophybalance}. The difference compared to the functions obtained from the DNS demonstrates, as expected, that pronounced correlations between the fields of the rate-of-strain tensor, the Laplacian of the vorticity, and the vorticity, respectively, exist.\par
To now quantify the contributions of the coherent structures, we start with investigating the diffusive term, which is presented in figure \ref{fig:diffusive}. It is observed that this term is almost fully represented by the coherent part of the field, while the incoherent contribution appears significantly smaller. As the Laplacian of a field enhances its small-scale features, this demonstrates that the CVE captures these features especially well. This is also supplemented by the observation made for the Fourier decomposition. It can be seen that the high pass filtered component is smaller, but not negligible for low magnitudes of vorticity, which means that both fields contribute to the enstrophy dissipation. To make this more quantitative, these contributions have been calculated and are presented in table \ref{tab:enstrophybudget}. It can be seen there that the low pass filtered component contributes only about $76 \%$ to the total dissipation of enstrophy compared to $91 \%$ in the case of the coherent vorticity.\par
Similar observations can be made for the terms related to the conditional vortex stretching term, which are shown in figure \ref{fig:vortexstretching}. Also for this term the coherent contribution matches almost perfectly the total contribution, about $98 \%$ of the enstrophy production is contained within this term (see table \ref{tab:enstrophybudget}). The remaining terms are strongly reduced in amplitude, still an investigation of their comparably small contributions is interesting, as can be seen in figure \ref{fig:vortexstretchingincoherent}. It becomes apparent from this figure that the interaction of the coherent part of the rate-of-strain field with the incoherent part of the vorticity field is positively correlated with the vorticity, i.e., a positive contribution to the average enstrophy budget originates from this term. An interesting interpretation of this observation is that the rate-of-strain field produced by the coherent vortex structures is able to produce additional coherent vortex structures; in a sense coherent structures breed coherent structures. In contrast to that, the interaction of the rate-of-strain field induced by the incoherent vorticity with the coherent vorticity has a depleting effect, such that it can be concluded that the incoherent rate-of-strain field destroys coherent vortex structures and hence has a dissipative character. The contribution of the incoherent rate-of-strain tensor times the incoherent vorticity field is negligible in view of its very low amplitude.\par
For the Fourier filtering it can be seen that the low pass filtered contribution of the rate-of-strain field times the low pass filtered contribution of the vorticity field does not fully represent the total vortex stretching term. However, still about $93 \%$ of enstrophy production is contained within this term (see table \ref{tab:enstrophybudget}). Although distinct differences compared to the CVE are apparent, table \ref{tab:enstrophybudget} shows that these differences almost vanish in the average. This exemplifies that more detailed insights can be obtained by studying conditional averages instead of ordinary averages. For the cross-terms similar observations can be made as in the case of the wavelet analysis. The contribution of the low pass filtered rate-of-strain field and the high pass filtered vorticity field is positively correlated with the vorticity, whereas the opposite is observed for the case of the high pass filtered rate-of-strain tensor contribution and the low pass filtered vorticity field. Apart from the comparison to the wavelet filtered data, this is a physically interesting observation on its own as it quantifies the interaction of large- and small-scale contributions within the flow fields. The contributions involving high pass filtered fields also remain small in the case of the Fourier-filtered data. Still, they are significantly larger than the corresponding terms of the wavelet decomposition; the scale in figure \ref{fig:vortexstretchingincoherent} differs by more than an order of magnitude between the wavelet and Fourier decomposition.

\section{Conclusion}
To summarize, we presented a detailed analysis of the conditional vorticity budget in terms of coherent vorticity. For this purpose we made use of the CVE technique to separate the noisy incoherent contributions of the vorticity field from the coherent ones. It was shown, in accordance with previous results, that CVE yields an excellent representation of the total flow using a strongly reduced number of degrees of freedom. This is particularly interesting as the conditional budget of vortex stretching and vorticity diffusion represents a dynamical rather than a purely kinematic relation. To further quantify the performance of the wavelet filtering method, we have performed a comparison to an ideal Fourier filter, where the fields have been low and high pass filtered.\par
Although visualizations did not show strong discrepancies between Fourier and wavelet filtering, the investigation of the conditional averages revealed more pronounced differences. It has been shown that most of the enstrophy production can be accounted to the coherent vorticity and the correspondingly induced rate-of-strain field. Interestingly, we have found that the incoherent rate-of-strain field tends to deplete vorticity, i.e., it tends to destroy coherent structures, while the rate-of-strain field induced by the coherent vorticity contributes positively. However, these contributions are small compared to the coherent-coherent contribution. In this sense the coherent structures are able to maintain or even amplify themselves, whereas the incoherent contributions tend to have a dissipative effect. Our analysis hence also quantifies the interaction of coherent and incoherent contributions to the flow. Consistent results have been found for the Fourier filtering. It has been shown, however, that this kind of filtering does not separate coherent contributions as well as wavelet filtering does, which is in agreement with previous studies.\par
These findings motivate to further develop CVS, where neglecting the incoherent flow contributions is assumed to be sufficient to model turbulent dissipation. With respect to the performance of this method, it should be noted that Coherent Vorticity Simulations using adaptive wavelet bases require, however, the use of a safety zone to account for the generation of wavelet coefficients in scale and space due to the nonlinear flow dynamics. This increases the number of retained coefficients in actual simulations by a factor 2-8 depending on the choice of the safety zone. For a discussion on possible choices and their influence on the flow statistics we refer the reader to a recent work \cite{okamoto11mms}. The Lagrangian particle-wavelet method \cite{bergdorf06mms} would allow to perform CVS without safety zone. Note that in simulations using Fourier spectral methods the number of modes has also to be increased by a certain factor in each spatial direction depending on the used dealiasing technique to remove high wavenumber modes for computing the nonlinear term.

\section*{Acknowledgments}
We would like to acknowledge the Math and Iter 2009 program and the CEMRACS 2010 summer program both at CIRM Luminy, where parts of this work have been carried out. Computational resources were granted within the project h0963 at the LRZ Munich. MF and KS acknowledge financial support from the PEPS program of INSMI-CNRS.

\end{document}